\documentstyle[12pt,fleqn,leqno]{article}

\begin{document}

\title{Exact Solution of a Boundary Value Problem in
Semiconductor Kinetic Theory}
\author{C.~Dalitz, E~H.~de Groot\\
Universit\"at Bielefeld, Theoretische Physik\\
Universit\"atsstr.~25, 33615 Bielefeld, Germany\\[1cm]}
\date{}
\maketitle
\begin{center}
\section*{Abstract}
\end{center}
\begin{it}
An explicit solution of the stationary one dimensional half-space
boundary value problem for the linear Boltzmann equation is presented
in the presence of an arbitrarily high constant external field. The
collision kernel is assumed to be separable, which is also known as
``relaxation time approximation''; the relaxation time may depend
on the electron velocity. Our method consists in a transformation of
the half-space problem into a nonnormal singular integral equation, 
which has an explicit solution. \\[5mm]
\end{it}

%
%
\newcommand{\R}{\mbox{{\rm I\hspace{-0.6mm}R}}}
\newcommand{\C}{{\rm \rlap{C}\hspace{-0.2mm}
                 \hbox{ \vrule width 0.6pt height 8pt depth 0pt 
                 \hskip 0pt}}}
%
%
\renewcommand{\thesection}{\Roman{section}}
\renewcommand{\thesubsection}{\Roman{section}.\Roman{subsection}}

\section{Introduction}
The motion of electrons or holes in a bulk semiconductor under
the action of an external field $E$ can be described by a probability
density $f(t,x,p)$ for finding a charge carrier at time $t$ and point $x$
with momentum $p$. Under the assumptions of quasineutrality \cite{Hout}
and a low electron concentration, so that interactions among the electrons
themselves (via collisions or {\sc Pauli}'s principle) can be neglected 
\cite{Markowich}, this probability density satisfies the linear Boltzmann 
equation
\begin{displaymath}
\Big( \partial_t +p\partial_x - E\partial_p \Big)f(t,x,p) ={\cal Q}f(t,x,p)
\end{displaymath}
where we have set the charge and effective mass of the electrons equal to 
unity. The linear collision operator ${\cal Q}$ describes scattering of the 
electrons by vibrations of the host lattice and consequently strongly 
depends on the model for the electron-phonon interaction. \par
In this article we consider collision models with separable collision
kernels. These models are also known as {\it relaxation time approximations}
with momentum dependent relaxation times. Although these models do not account 
for some high field effects in semiconductors \cite{Dalitz_moments}, they are 
widely used because of their relatively simple mathematical structure.
Whilst the solution of the initial
value problem is easily obtained in these models \cite{Arlotti} \cite{Tang},
the boundary value problem is far more intricate. Up to now, a solution of
the boundary value problem in the relaxation time approximation has been 
found only by {\sc Cercignani} for a specific $p$-dependence of the 
relaxation time \cite{Cercignani}
and by {\sc Dalitz} in the zero temperature limit \cite{Dalitz_zero}. \par
In this article, we present a solution of the stationary one dimensional
half-space problem with a constant electric field for quite general 
$p$-dependent relaxation times. The somewhat simpler case of a constant
relaxation time is treated in the unpublished preprint \cite{deGroot}
in great detail. In the half-space problem the semiconductor is taken 
semiinfinite ($x\in(0,\infty)$); at the boundary $x=0$, electrons are shot 
in with a given momentum distribution $\varphi$. Thus we seek the solution of
\begin{equation}
\Big(p\partial_x - E\partial_p\Big) f(x,p) = {\cal Q}f(x,p)
\end{equation}
with the boundary conditions 
\begin{equation}
f(0,p)=\varphi(p)\quad\mbox{for}\quad p>0 \quad\mbox{ and }\quad 
\lim_{x\to\infty}f(x,p)=0
\end{equation}
\indent
For zero electric field $E\equiv 0$ {\sc Case} \cite{Case} has found a solution 
via an eigenfunction technique in the context of neutron diffusion in solids. 
Completeness theorems for these eigenfunctions have been proven by {\sc Zweifel}
\cite{Zweifel} for the general equation $h(p)\partial_xf+{\cal A}f=0$ under the 
assumption that ${\cal A}$ be a selfadjoint positive operator in some appropriate 
function space. In the case of a nonzero electric field however, this general 
result is no longer valid because the operator ${\cal A}=-E\partial_p-{\cal Q}$ 
is not selfadjoint.\par
Nevertheless for specific collision operators, an eigenfunction expansion of the
solution might be possible also in the presence of a constant external field. 
In \cite{Dalitz_exact} a kind of half-range completeness of the eigenfunctions
has been proven for a constant relaxation time in the zero temperature limit, 
and in \cite{Dalitz_half} the same has been proven for two very specific
collision kernels at arbitrary temperature of the semiconductor host lattice.
In this article we show that the half-space problem in the relaxation time
approximation generally can be solved via an eigenfunction expansion, provided
the collision frequency is an analytic function of the momentum $p$ that
increases not more than linearly in $p$ for $p\to\infty$. \par
We shall prove that each boundary value $\varphi$ that is a Laplace transform 
of any tempered distribution can be written as a superposition of 
``eigenfunctions''. This result (see end of section V) is in complete 
analogy to {\sc Case}'s half-range completeness theorem for his ``singular 
eigenfunctions''. Moreover we shall see that the eigenfunction representation
of the boundary value generally contains a singular contribution of the 
Maxwellian of the temperature of the semiconductor lattice. This is in 
agreement with a conjecture of {\sc Stichel} and {\sc Strothmann} \cite{Stichel}.

\section{The collision model}
In the nondegenerate situation, the electron phonon collision
operator ${\cal Q}$ in (1) has the general form 
\cite{Markowich} 
\begin{equation}
{\cal Q}f(p) = M(p)\int dp'\,K(p',p)\,f(p') - f(p)\int dp'\,M(p')\,K(p,p')
\end{equation}
where $M(p):=M_0\,e^{-p^2\!/2\theta}$ is the Maxwellian 
at temperature $\theta$ of the semiconductor lattice, which is assumed to be
in equilibrium. The {\it collision
kernel} $K(p,p')$ is a symmetric and positive distribution. The second integral
in (3), which gives the total scattering rate, is called {\it collision
frequency} $\nu$; its reciprocal is called {\it relaxation time} $\tau$
\begin{equation}
\nu(p)=\frac{1}{\tau(p)}:=\int dp'\,M(p')\,K(p,p')
\end{equation}
Throughout this article we assume the collision kernels to be seperable, that
is $K(p,p')=\nu(p)\cdot\nu(p)=(\tau(p)\cdot\tau(p'))^{-1}$. This assumption
is also known as {\it relaxation time approximation} (RTA). If we
normalize the Maxwellian so that $\int M(p)\,\nu(p)\,dp=1$, we obtain the 
RTA in its usual form
\begin{equation}
{\cal Q}f(p) = \frac{1}{\tau(p)}
\left(M(p)\,\int\!d^3p'\,\frac{f(p')}{\tau(p')}-f(p)\right)
\end{equation}
and by comparison with (4) we see that $\tau(p)$ indeed is the relaxation time, 
because of our normalization $\int M/\tau=1$. The relaxation time approximation 
(5) generally can be a good approximation for the collision operator, provided 
the collision kernel $K$ is a measurable function with a finite norm
$\int dp\int dp'\,M(p)\,M(p')\,K^2(p,p')<\infty$. For then
(5) is equivalent to keeping only the largest term in the eigenfunction 
representation of the symmetric and square integrable kernel 
$\sqrt{M(p)}K(p,p')\sqrt{M(p')}$ \cite{remark1}. 

\section{Eigenfunction expansion of the solution}
In order to solve the stationary Boltzmann equation (1), we make the
separation ansatz $f_{\lambda}(x,p):=e^{-\lambda x}g_{\lambda}(p)$.
We then obtain for the functions $g_{\lambda}$ the "eigenvalue" 
equation $\left( {\cal Q}+E\partial_p \right) g_{\lambda}(p) = -\lambda p\,
g_{\lambda}(p)$, which reads with the collision operator (5)
\begin{equation}
\Big(E\partial_p + \lambda p-\nu(p)\Big)g_\lambda(p)=
-\nu(p)\,M(p)\int\!dp'\,\nu(p')\,g_\lambda(p')
\end{equation}
In order to get rid of the integral on the right hand side of (6), let us
normalize $\int\nu\,g_\lambda=1$. Then the general solution of (6) reads
\begin{equation}
g_\lambda(p) = \frac{M_0}{E}\;e^{-\lambda p^2\!/2E + N(p)/E}
\left\{ C_\lambda - \int\limits_0^p\!dq\,\nu(q)\,
e^{(\lambda-E/\theta) q^2\!/2E - N(q)/E} \right\}
\end{equation}
where $C_\lambda$ is an integration constant and $N$ (read ``capital $\nu$'') 
denotes the primitive function of the collision
frequency
\begin{equation}
N(p):=\int_0^p\!dq\,\nu(q)
\end{equation}
For general $\lambda$, (7) is integrable with respect to $p$ only if 
$\lim_{p\to\infty}\nu(p)/p=c<\infty$. In fact, if
this condition is violated, the only $L_1$-integrable solutions of (6) are
$M(p)$ with $\lambda=E/\theta$ and the homogeneous solution with 
$\lambda=0$. If the condition is satisfied, the solution (7) is 
$L_1$-integrable for every $\lambda\cdot\mbox{sign}(E)>c$. For simplicity 
we assume that $c=0$. \par
Because of the boundary condition at infinity (2), only positive values
are allowed for $\lambda$. Consequently, we must assume that $E>0$, for
otherwise $g_\lambda$ would grow exponentially for $p\to\pm\infty$. 
This means that the field $E$ must act in such a way that the electrons
are driven back to the boundary $x=0$.\par
Inserting the eigenfunction (7) into the normalization condition 
$\int\!dp\,\nu\,g_\lambda=1$, we find for the constant $C_\lambda$ after 
a partial integration in the numerator
\begin{equation}
C_\lambda = \frac{\int\limits_{-\infty}^\infty\!dp\,p\,e^{-\lambda p^2\!/2E+N(p)/E}
\int\limits_0^p dq\,\nu(q)\,e^{(\lambda-E/\theta)q^2\!/2E - N(q)/E}}
{\int\limits_{-\infty}^\infty\!dp\,p\,e^{-\lambda p^2\!/2E+N(p)/E}}
\end{equation}
Since equation (1) is linear, any superposition of solutions $f_{\lambda}(x,p)$ is
a solution of (1) too. Thus we assume the solution to be of the form
\begin{equation}
f(x,p)=\int_0^\infty\!d\lambda\,A(\lambda)\,f_{\lambda}(x,p)
=\int_0^\infty\!d\lambda\,A(\lambda)\,e^{-\lambda x}\,g_{\lambda}(p)
\end{equation}
The expansion coefficients $A(\lambda)$ must be determined from the boundary 
value $\varphi$. Setting $x\!=\!0$ in (10) yields
\begin{equation}
f(0,p)=\varphi(p)=\int_0^\infty\!d\lambda\,A(\lambda)\,g_{\lambda}(p)
\quad\mbox{for}\quad p>0
\end{equation}
If this equation can be solved for $A(\lambda)$, then the half-space problem
(1) and (2) is solved by (10). Hence the task is to determine the class of 
boundary values $\varphi$ for which (11) has a solution $A(\lambda)$ and to
determine this solution in terms of $\varphi$. We shall see that (11) indeed
can be solved if the boundary value is a Laplace transform of any tempered
distribution, for then (11) is equivalent to a singular integral equation
which can be solved explicitly. \par
It is interesting to note that the representation (10) of the solution of the
stationary Boltzmann equation (1) can
be written in a different form which is more general than the representation
via eigenfunctions. If we define
\begin{displaymath}
h(\xi):=\int\limits_0^\infty\!d\lambda\,e^{-\lambda\xi/2E}A(\lambda)
\left(\int_{-\infty}^\infty\!dp\,\nu(p)\,e^{-\lambda p^2\!/2E+N(p)/E}\right)^{-1}
\end{displaymath}
then $f(x,p)=\int_0^\infty\!d\lambda\,A(\lambda)\,e^{-\lambda x}g_\lambda(p)$
takes the form
\begin{eqnarray*}
f(x,p) & = & e^{N(p)/E}\,h(p^2+2Ex) + \frac{e^{N(p)/E}}{E}
             \int\limits_{-\infty}^\infty\!dp'\,\nu(p')\,e^{N(p')/E} \times \\
 & & \int\limits_p^{p'}\!dq\,\nu(q)\,e^{-N(q)/E}M(q)\,h(p^2+{p'}^2-q^2+2Ex) 
\end{eqnarray*}
The boundary condition at $x=0$ leads to an integral equation for the
unknown function $h$. Maybe a solution of this equation is possible for a
more general class of boundary values $\varphi$.

\section{A singular integral equation for the expansion coefficients}
If we insert the eigenfunctions (7) into the boundary 
condition $f(0,p)=\varphi(p)$ for $p\!>\!0$ and make the substitutions
$t\!=\!p^2\!/2E$ and $s\!=\!q^2\!/2E$, the integral equation (11) for the
expansion coefficients $A(\lambda)$ reads
\begin{equation}
\frac{E}{M_0}\;e^{-N(\sqrt{2Et})/E}\varphi(\sqrt{2Et})\quad=\quad
\int\limits_0^\infty\!d\lambda\,A(\lambda)\,C_\lambda\,e^{-\lambda t}
\end{equation}
\vspace{-5mm}
\begin{displaymath}
-\,\int\limits_0^\infty\!d\lambda\,A(\lambda)\,e^{-\lambda t}
\int\limits_0^t\!ds\,e^{\lambda s}\sqrt{E/2s}\;
\nu(\sqrt{2Es})\,e^{-N(\sqrt{2Es})-sE/\theta}
\end{displaymath}
Now let us make the assumption that the left hand side of (12) can be written
as a Laplace transform of any tempered distribution $\mit\Psi$
\begin{equation}
\frac{E}{M_0}\;e^{-N(\sqrt{2Et})/E}\varphi(\sqrt{2Et}) = 
\int_0^\infty\!d\lambda\,e^{-\lambda t}\mit\Psi(\lambda)
\end{equation}
Additionally let us assume that there is a continuous function ${\cal N}$ 
with ${\cal N}(\mu)\!>\!0$ for $\mu\!>\!0$ and
\begin{equation}
\sqrt{E/2s}\;e^{-N(\sqrt{2Es})/E}\nu(\sqrt{2Es})
=\int_0^\infty\!d\mu\,e^{-\mu s}{\cal N}(\mu)
\end{equation}
Examples for which this holds are $\nu(p)=a p^\alpha$ with 
$-1\!<\!\alpha\!\le 1$ and sums of these powers. If we integrate (14) from
$s$ to infinity, we find
\begin{equation}
E\,e^{-N(\sqrt{2Es})/E} = \int_0^\infty\!d\mu\,e^{-\mu s}{\cal N}(\mu)/\mu
\end{equation}
Therefore ${\cal N}/\mu$ is locally integrable at $\mu\!=\!0$ and consequently
${\cal N}(0)\!=\!0$. Integrating a second time from $s$ to infinity, we find
$\lim_{\mu\to 0}{\cal N}/\mu\!=\!0$, so that $\beta(\lambda)$ defined below 
(see 19) is a continuous function of $\lambda$.
Insertion of (13) and (14) into (12) and usage of the shifting theorem
of the Laplace transform leads to (remember uniqueness of the Laplace transform)
\begin{eqnarray}
\mit\Psi(\lambda) & = & \left\{ C_\lambda + {\cal P}\!\int\limits_0^\infty\!
d\mu\,\frac{{\cal N}(\mu)}{\lambda -E/\theta -\mu} \right\} A(\lambda) \\
 & & -\;\Theta(\lambda -E/\theta)\,{\cal N}(\lambda -E/\theta)\;
{\cal P}\!\int\limits_0^\infty\!d\mu\,\frac{A(\mu)}{\mu-\lambda} \nonumber
\end{eqnarray}
where the symbol ``${\cal P}$'' means Cauchy's principal value and $\Theta$
denotes Heaviside's step function. This equation is a singular integral equation
for the expansion coefficients $A(\lambda)$. In the limit $\theta\to 0$, the
$\Theta$-function in front of the 
singular integral over $A$ vanishes, which simplifies the situation considerably
and allows the treatment described in \cite{Dalitz_zero}, section 6.\par
In contrast to ``normal'' singular equations however, the right hand side of 
(16) has a zero at $\lambda\!=\!E/\theta$. Thus we are dealing with a 
``nonnormal'' problem. This zero is easily split off by a partial integration 
using $-E\partial_q e^{-N/E}\!=\!\nu e^{-N/E}$ in the numerator of 
$C_\lambda$ (see 9) and usage of $E\!=\!\int_0^\infty\!d\mu\,{\cal N}/\mu$
(see 15). The result is
\begin{equation}
\mit\Psi(\lambda) = \Big(\lambda-E/\theta\Big)\cdot\left\{ \alpha(\lambda)\,
A(\lambda)+ \beta(\lambda)\;{\cal P}\!\int\limits_0^\infty\!d\mu\,
\frac{A(\mu)}{\mu-\lambda} \right\}
\end{equation}
with the notations
\begin{eqnarray}
\alpha(\lambda) & = & {\cal P}\!\int\limits_0^\infty\!d\mu\,
\frac{{\cal N}(\mu)}{\mu(\lambda-\mu-E/\theta)} \\
& & + \; \frac{\int\limits_{-\infty}^\infty\!dp\,p\,e^{-\lambda p^2\!/2E+N(p)/E}
\int\limits_0^p dq\,q\,e^{(\lambda-E/\theta)q^2\!/2E - N(q)/E}}
{\int\limits_{-\infty}^\infty\!dp\,p\,e^{-\lambda p^2\!/2E+N(p)/E}} 
\nonumber \\[5mm]
\beta(\lambda) & = & -\Theta(\lambda-E/\theta)\;
\frac{{\cal N}(\lambda -E/\theta)}{\lambda-E/\theta}
\end{eqnarray}

\section{Solution of the singular integral equation}
In appendix A.I we will prove that $\alpha(\lambda)<0$ for $\lambda\le E/\theta$.
Therefore $\alpha(\lambda)\pm i\pi\beta(\lambda)$ has no zeroes for $\lambda\in\R^+$, 
and we can transform the nonnormal problem (17) into a normal problem simply by 
dividing both sides by $(\lambda-E\theta)$. \par
However, if we divide the left
hand side of (17) by $(\lambda-E/\theta)$, we obtain a {\it distribution}, even
if $\mit\Psi$ is a smooth function. Although the standard theory of singular
integral equations \cite{Muskhelishvili} only deals with H\"older-continuous
functions, physicists have for a long time been applying the same methods to 
distributions as well (see \cite{Inonu} for a comprehensive but heuristic
treatment); meanwhile {\sc Estrada} and {\sc Kanwal} have presented a rigorous 
theory \cite{Estrada}. Recently the semiinfinite Hilbert transform of distributions
has been discussed in a more general framework \cite{Baptiste}. \par
If we divide (17) by $(\lambda-E/\theta)$, we obtain the normal singular integral 
equation
\begin{equation}
\mit\Upsilon(\lambda) = \alpha(\lambda)\,A(\lambda)+
\beta(\lambda)\;{\cal P}\!\int\limits_0^\infty\!d\mu\;\frac{A(\mu)}{\mu-\lambda}
\end{equation}
The tempered distribution $\mit\Upsilon$ is the most general solution
of the (distributional) equation $\mit\Upsilon\cdot(\lambda-E/\theta)=\mit\Psi$.
Thus we can write 
\begin{equation}
\mit\Upsilon(\lambda)={\cal P}\frac{\mit\Psi(\lambda)}{\lambda-E/\theta}
+ c_E\,\delta(\lambda-E/\theta)
\end{equation}
where $c_E$ is an arbitrary constant which will be essential later on.
The symbol ``${\cal P}$'' stands for
``particular solution'', which is known to exist from {\sc H\"ormander}'s theorem
\cite{Hoermander}. If we define principal value integrals in the more
general sense of \cite{Estrada} or \cite{Baptiste}, we may read this symbol as 
``principal value''. \par
The zero on the right hand side of (17) has introduced an arbitrary constant
into our problem. This is in agreement with {\sc Pr\"ossdorf}'s result in a
nondistributional framework \cite{Proessdorf} that such a zero increases the 
index of the equation by one. \par
Although the problem of ``nonnormality'' is solved, there is still a
problem with eq.~(20). The standard theory of singular integral equations 
is only applicable to equations on a {\it finite} interval, but our equation
lives on a semiinfinite interval. We can circumvent this difficulty with the
transformation
\begin{equation}
\mu=\frac{1+y}{1-y} \quad\quad\mbox{ and }\quad\quad \lambda=\frac{1+x}{1-x}
\end{equation}
which maps the interval $0\!<\!\lambda\!<\!\infty$ monotonously onto the interval
$-1\!<\!x\!<\!1$. This transformation does not seem to be well known. For instance 
{\sc Paveri-Fontana} and {\sc Zweifel} \cite{Paveri+Zweifel} recently presented an
ab initio derivation of an inversion formula for the half-Hilbert transform which 
turns out to be equivalent to the well known inversion formula on a finite
interval \cite{Tricomi} if the transformation (22) is made. \par
With this transformation, equation (20) transforms into
\begin{equation}
\hat{\mit\Upsilon}(x) = \hat{\alpha}(x)\,\hat{A}(x)+\hat{\beta}(x)\;{\cal P}\!
\int\limits_{-1}^1dy\;\frac{\hat{A}(y)}{y-x}
\end{equation}
with the definitions
\begin{eqnarray*}
\hat{\mit\Upsilon}(x):=\frac{\mit\Upsilon\left(\displaystyle\frac{\displaystyle 1+x}
{\displaystyle 1-x}\right)}{1-x} & \quad \mbox{and} \quad &
\hat{A}(x):=\frac{A\left(\displaystyle
\frac{\displaystyle 1+x}{\displaystyle 1-x}\right)}{1-x} \\[3mm]
\hat{\alpha}(x):=\alpha\left(\frac{1+x}{1-x}\right) & \quad \mbox{and} \quad &
\hat{\beta}(x):=\beta\left(\frac{1+x}{1-x}\right) 
\end{eqnarray*}
Although the theory of singular integral equations on a finite interval is well
developed (see \cite{Inonu} or \cite{Myschkis} for a comprehensive summary), we will 
sketch the solution of (23) in some detail. Let us start with the definition of the 
analytic function $F$ associated with the distribution $\hat{A}$:
\begin{displaymath}
F(z):=\frac{1}{2\pi i}\int\limits_{-1}^1 dx'\;\frac{\hat{A}(x')}{x'-z}
\quad\mbox{ for }z\in\C\;\setminus [-1,1]
\end{displaymath}
which is a holomorphic function in $\C\;\setminus[-1,1]$. If $z$ approaches the cut
$[-1,1]$ from above or below, its boundary values are given by the Plemelj formula
\begin{displaymath}
\lim_{\epsilon\to 0}F(x\pm i\epsilon)=F^\pm(x)=
\frac{1}{2\pi i}\,{\cal P}\!\int\limits_{-1}^1dx'\;\frac{\hat{A}(x')}{x'-x}
\pm\frac{1}{2}\hat{A}(x)
\end{displaymath}
Hence our integral equation (23) can be written as a relation between these two
boundary values (remember that $\hat{\alpha}\pm \pi i\hat{\beta}\neq 0$)
\begin{equation}
\frac{\hat{\mit\Upsilon}(x)}{\hat{\alpha}(x)-\pi i\hat{\beta}(x)}=
\frac{\hat{\alpha}(x)+\pi i\hat{\beta}(x)}{\hat{\alpha}(x)-\pi i\hat{\beta}(x)}
F^+(x) - F^-(x)
\end{equation}
which is known as a {\it Riemann-Hilbert problem} in the literature. 
Now we are looking for a function $\chi$ which is holomorphic in the complex plane
cut at $[-1,1]$ and satisfies the boundary condition
\begin{displaymath}
\frac{\hat{\alpha}(x)+\pi i\hat{\beta}(x)} 
{\hat{\alpha}(x)-\pi i\hat{\beta}(x)} = \frac{\chi^+(x)}{\chi^-(x)}
\end{displaymath}
If we find such a function, we can determine $\chi F$ from (24) and therefore also
$F$, and from $\hat{A}=F^+-F^-$ we can obtain $\hat{A}$. A function with
this property is
\begin{eqnarray}
\chi(z) & = & (z+1)^m (z-1)^n e^{\mit\Gamma(z)} \quad\mbox{ with} \\
\mit\Gamma(z) & := & \frac{1}{\pi}\int\limits_{-1}^1dx'\;\frac{\gamma(x')}{x'-z} 
\quad\mbox{ and} \\
\gamma(x) & := & \frac{1}{2i}\ln\left(
\frac{\hat{\alpha}+\pi i\hat{\beta}}{\hat{\alpha}-\pi i\hat{\beta}}\right)
=\mbox{arccot}\left(\frac{\hat{\alpha}(x)}{\pi\hat{\beta}(x)}\right)
\end{eqnarray}
In (27) we are choosing the main branch of the arcus cotangens (see appendix A.II).
The exponents $m$ and $n$ must be chosen in such a way that $\chi$ and $F$ are 
integrable near the end points $z=\pm 1$ (see \cite{Myschkis} for details; this 
is the crucial point where the interval must be finite):
\begin{eqnarray*}
-1 < & m - \lim_{x\to-1}\gamma(x)/\pi & < 1\\
-1 < & n + \lim_{x\to+1}\gamma(x)/\pi & < 1
\end{eqnarray*}
In appendix A.II it is proven that $\gamma(-1)=0$ and $\gamma(+1)=\pi$. Therefore
we must choose $m=0$ and $n=-1$. Thus the {\it index} of the integral equation,
which is defined by $m+n$, is $-1$. Consequently, (23) has a solution $\hat{A}$ only 
if the inhomogeneity $\hat{\mit\Upsilon}$ satisfies the {\it orthogonality relation}
\cite{Inonu}
\begin{equation}
\int\limits_{-1}^1dx\;\frac{\hat{\mit\Upsilon}(x)\chi^-(x)}
{\hat{\alpha}(x)-\pi i\hat{\beta}(x)} = 0
\end{equation}
Fortunately $\mit\Upsilon$ contains an arbitrary constant, which we will choose
in such a way that this orthogonality relation holds.
If (28) is satisfied, the solution of our integral equation (23) is unique and
reads
\begin{eqnarray*}
\hat{A}(x) & = & F^+(x) - F^-(x) \quad\mbox{ with} \\
F(z) & = & \frac{1}{\chi(z)}\frac{1}{2\pi i}\int\limits_{-1}^1 dx'\;
\frac{\chi^-(x')\,\hat{\mit\Upsilon}(x')}{\left[\hat{\alpha}(x')-
\pi i\hat{\beta}(x')\right]\,(x'-z)}
\end{eqnarray*}
By application of the Plemelj formula to (25)-(27) we can express $\chi^\pm$ in
terms of principal value integrals. Moreover, the solution of the original 
equation on the semiinfinite interval $(0,\infty)$ is easily obtained via the 
transformation inverse to (22). If we set 
$x=(\lambda-1)/(\lambda+1)$ and $x'=(\mu-1)/(\mu+1)$, we obtain 
\begin{equation}
A(\lambda) = \frac{\alpha(\lambda)}{\alpha^2(\lambda)+\pi^2\beta^2(\lambda)}\;
\mit\Upsilon(\lambda)
\end{equation}
\begin{displaymath}
-\;\frac{\beta(\lambda)\,e^{-G(\lambda)}}
{\sqrt{\alpha^2(\lambda)+\pi^2\beta^2(\lambda)}}\;
{\cal P}\!\int\limits_0^\infty\!d\mu\;\frac{(\mu+1)\,e^{G(\mu)}\mit\Upsilon(\mu)}
{(\lambda+1)\sqrt{\alpha^2(\mu)+\pi^2\beta^2(\mu)}\,(\mu-\lambda)}
\end{displaymath}
with the notations
\begin{eqnarray}
G(\lambda) & := & \frac{\lambda+1}{\pi}\;{\cal P}\!
\int\limits_0^\infty\!d\mu\;\frac{g(\mu)}{(\mu+1)(\mu-\lambda)}\quad\mbox{ and} \\
g(\mu) & := & \frac{1}{2i}\ln\left(\frac{\alpha+\pi i\beta}{\alpha-\pi i\beta}\right)
= \mbox{arccot}\left( \frac{\alpha(\mu)}{\pi\beta(\mu)} \right)
\end{eqnarray}
The orthogonality relation for the inhomogeneity $\mit\Upsilon$ becomes
\begin{equation}
\int\limits_0^\infty\!d\mu\;\frac{\mit\Upsilon(\mu)\,e^{G(\mu)}}
{\sqrt{\alpha^2(\mu)+\pi^2\beta^2(\mu)}} = 0
\end{equation}
According to {\sc Noether}'s theorem on singular integral equations \cite{Noether}, 
a necessary and sufficient condition for the solvability of the inhomogeneous
equation (20) is the orthogonality of the inhomogeneity $\mit\Upsilon$
to all solutions $B(\lambda)$ of the adjoint homogeneous equation
\begin{equation}
\alpha(\lambda)\,B(\lambda) - {\cal P}\!\int\limits_0^\infty\!d\mu\;
\frac{\beta(\mu)\,B(\mu)}{\mu-\lambda} = 0
\end{equation}
In our situation, this equation only has one solution. Comparison of {\sc Noether}'s
orthogonality condition $\int \mit\Upsilon B =0$ with (32) shows (the 
same result is obtained, of course, by direct solution of (33)) 
\begin{equation}
B(\lambda) = \frac{e^{G(\lambda)}}{\sqrt{\alpha^2(\lambda)+\pi^2\beta^2(\lambda)}}
\end{equation}
Because of $\frac{\mu+1}{(\mu-\lambda)(\lambda+1)}=\frac{1}{\mu-\lambda}+
\frac{1}{\lambda+1}$ and (32), we may write (29) as
\begin{displaymath}
A(\lambda)  =  \frac{1}{\alpha^2(\lambda)+\pi^2\beta^2(\lambda)}
\left\{ \alpha(\lambda)\,\mit\Upsilon(\lambda) 
 -\;\frac{\beta(\lambda)}{B(\lambda)}\;{\cal P}\!
\int\limits_0^\infty\!d\mu\;\frac{B(\mu)\,\mit\Upsilon(\mu)}{\mu-\lambda} \right\}
\end{displaymath}
Now let us insert $\mit\Upsilon$ according to (21). The constant $c_E$ must be
chosen
\begin{equation}
c_E = -\frac{1}{B(E/\theta)}\;\int\limits_0^\infty\!d\mu\,B(\mu)\,
{\cal P}\frac{\mit\Psi(\mu)}{\mu-E/\theta}
\end{equation}
so that the orthogonality relation (32) is satisfied. Finally we arrive at the
solution of the singular integral equation for the expansion coefficients:
\begin{equation}
A(\lambda) = \left. \frac{1}{\alpha^2(\lambda)+\pi^2\beta^2(\lambda)} \right\{
\alpha(\lambda)\,{\cal P}\frac{\mit\Psi(\lambda)}{\lambda-E/\theta}
\end{equation}
\begin{displaymath}
+\;\alpha(\lambda)\,c_E\,\delta(\lambda-E/\theta) -
\frac{\beta(\lambda)}{B(\lambda)\,(\lambda-E/\theta)}\; \left.
{\cal P}\!\int\limits_0^\infty\!d\mu\;\frac{B(\mu)\,\mit\Psi(\mu)}{\mu-\lambda}
\right\}
\end{displaymath}
where $B(\lambda)$ is given by (34). \par
Obviously, our solution 
$f(x,p)=\int d\lambda\,A(\lambda)\,e^{-\lambda x}g_{\lambda}(p)$ is well 
defined only if $\mit\Psi /\alpha$ is locally integrable at $\lambda=0$. 
From eq.~(A.1) in the appendix we conclude that for $\lambda\to 0$
\begin{displaymath}
1/\alpha(\lambda) \sim \mbox{const}\cdot \int_{-\infty}^\infty dp\,p\,
e^{-\lambda p^2\!/2E + N(p)/E} \to\infty
\end{displaymath}
Thus, if we demand
\begin{eqnarray}
\infty & > & \int\limits_0^\infty\!dp\,p\,\varphi(p) = 
\int\limits_0^\infty\!dp\,p\,\varphi(p)\,e^{-N(p)/E}\,e^{N(p)/E} \\
 & & \stackrel{(4.44)}{=} \frac{M_0}{E}\int\limits_0^\infty\!d\lambda\,
\mit\Psi(\lambda)\int\limits_0^\infty\!dp\,p\,e^{-\lambda p^2\!/2E + N(p)/E}
\nonumber
\end{eqnarray}
the solution of the half-space problem is given by the eigenfunction expansion
(10) with expansion coefficients (36). \par
Let us repeat this result in the form of a half-range completeness theorem for 
the eigenfunctions $g_\lambda$.\\[5mm]
{\bf Theorem:} 
\begin{it}
The integral $N$ over the collision frequency be such that
\begin{displaymath}
E\,e^{N(p)/E} = \int_0^\infty\!d\mu\,e^{\mu p^2\!/2E}{\cal N}(\mu)/\mu
\end{displaymath}
for some continuous function ${\cal N}$ with ${\cal N}(\mu)>0$ for $\mu>0$.
Then each function $\varphi:\R^+\to\R$ with $\int_0^\infty\!dp\,p\,\varphi<\infty$, 
that is a Laplace transform of any tempered distribution $\mit\Phi$
\begin{displaymath}
\varphi(p)=\int_0^\infty\!d\mu\,{\mit\Phi}(\mu)\,e^{-\mu p^2/2E}
\end{displaymath}
is a superposition of eigenfunctions
\begin{displaymath}
\varphi(p)=\int_0^\infty\!d\lambda\,A(\lambda)\,g_\lambda(p)
\end{displaymath}
where the expansion coefficients are given by (36) with 
\begin{equation}
\mit\Psi(\lambda) = \frac{1}{M_0}\int_0^\lambda d\mu\;
\frac{{\cal N}(\mu)}{\mu}\;\mit\Phi(\lambda-\mu)
\end{equation}
\end{it}
{\bf Remarks:} a) Relation (38) between $\mit\Psi$ and $\mit\Phi$ is an immediate
consequence of (14) and the convolution theorem of the Laplace transform. \par
b) According to (36), the expansion coefficients $A(\lambda)$ generally contain
a singular contribution from the eigenvalue $\lambda=E/\theta$, the corresponding
eigenfunction of which is the Maxwellian $M(p)$ at temperature $\theta$. This 
confirms a conjecture of {\sc Stichel} and {\sc Strothmann} \cite{Stichel}, who
took this as a starting point for an asymptotic analysis of the boundary value
problem. 

\section{Examples}
\subsection{Maxwellian as boundary value}
If the boundary value is a Maxwellian of the temperature $\theta$ of the host 
medium, that is $\varphi(p)=M(p)=M_0\,e^{-p^2\!/2\theta}$, the solution of the 
half-space problem reads $f(x,p)=e^{-Ex/\theta}M(p)$, as can be deduced from 
(1) with the use of ${\cal Q}M\equiv 0$. Hence we should obtain
$A(\lambda)=\delta(\lambda-E/\theta)$, so that this situation provides a nice
test of our solution. \par
Indeed, for $\varphi=M$ the inverse Laplace transform $\mit\Psi$ defined in 
(13) reads
\begin{displaymath}
\mit\Psi(\lambda) = \Theta(\lambda-E/\theta)\,\frac{{\cal N}(\lambda-E/\theta)}
{\lambda-E/\theta} = -\beta(\lambda)
\end{displaymath}
Thus the expansion coefficients (36) read in this case
\begin{eqnarray*}
A(\lambda) & = & \left. \frac{1}{\alpha^2(\lambda)+\pi^2\beta^2(\lambda)} 
\right\{ \alpha(\lambda)\, c_E\delta(\lambda-E/\theta) \\
& & \left. -\;\frac{\alpha(\lambda)\,\beta(\lambda)}{\lambda-E/\theta}
+\frac{\beta(\lambda)}{B(\lambda)\,(\lambda-E/\theta)}\;
{\cal P}\!\int\limits_0^\infty\!d\mu\;
\frac{B(\mu)\,\beta(\mu)}{\mu-\lambda} \right\}
\end{eqnarray*}
and the constant $c_E$ is given by (note that the symbol ${\cal P}$ can be
omitted because of $\beta(\lambda)\equiv 0$ for $\lambda\le E/\theta$)
\begin{displaymath}
c_E = \frac{1}{B(E/\theta)}\int\limits_0^\infty\!d\mu\;\frac{B(\mu)\,\beta(\mu)}
{\mu -E/\theta}
\end{displaymath}
Because of relation (33), the integrals over $B\beta$ can be 
replaced by $\alpha(\lambda)\,B(\lambda)$ and $\alpha(E/\theta)\,B(E/\theta)$. 
If we take care of $\beta(E/\theta)=0$, we obtain indeed 
$A(\lambda)=\delta(\lambda-E/\theta)$.

\subsection{Constant relaxation time}
In case of a constant relaxation time $\nu(p)=const.=1/\tau$, the quantities
$\alpha$, $\beta$ and $B$ in the formula (36) for the expansion coefficients
can be calculated explicitly. Moreover there is a relation between $\alpha$
and $\beta$ that allows a simplification of some expressions. For details of 
the calculation we refer the interested reader to \cite{deGroot}.\par
Without loss of generality we can set $\tau$ and $\theta$ equal unity 
\cite{remark2}. Then the function ${\cal N}$ that is necessary for the calculation 
of $\mit\Psi$ via (38) reads ${\cal N}(\mu)=(E/2\pi\mu)^{1/2}e^{-1/2E\mu}$ and 
the functions $\alpha$ and $\beta$ are given by
\begin{eqnarray*}
\alpha(\lambda) & = & \sqrt{\frac{4E}{\pi}}\lambda^{3/2}e^{-1/2E\lambda}\;
                      {\cal P}\int\limits_0^\infty\!dx\,\frac{x^{-3/2}e^{-1/x}}
                      {2\lambda(\lambda-E)-x} \\
\beta(\lambda) & = & -\Theta(\lambda-E)\sqrt{\frac{E}{2\pi}}
                      \frac{e^{-1/2E(\lambda-E)}}{(\lambda-E)^{3/2}}
\end{eqnarray*}
where the principal value integral in $\alpha$ allows for an explicit 
evaluation
\begin{displaymath}
{\cal P}\int\limits_0^\infty\!dx\,\frac{x^{-3/2}e^{-1/x}}{y-x} = \left\{
\begin{array}{ll}
y\sqrt{\pi}\left( 1-\sqrt{-y}\,e^{-y}2\int_{\sqrt{-y}}^\infty e^{-t^2}\,dt 
          \right) & \mbox{ for }y\le 0 \\[3mm]
y\sqrt{\pi}\left( 1-2\sqrt{y}\,e^{-y}\int_0^{\sqrt{y}}e^{t^2}\,dt \right)
          & \mbox{ for }y\ge 0
\end{array}\right. 
\end{displaymath}
Moreover in this situation the function $B(\lambda)$, which generally 
contains a principal value integral (see (34) and (30)), can be expressed
in terms of a regular integral. We find (apart from an irrelevant constant 
which cancels out in (36))
\begin{eqnarray*}
B(\lambda) & \propto & \lambda^{-3/2}e^{1/2E\lambda} \xi(\lambda) 
                       \quad\mbox{ with} \\
\xi(\lambda) & = & \exp\left\{ \frac{E}{2\pi}\int\limits_E^\infty\!
                   d\mu\,\frac{g(\mu)}{\mu+\lambda-E}\;
                   \frac{2\lambda-E}{2\mu-E} \right\}
\end{eqnarray*}
where $g(\mu)=\mbox{arccot}(\alpha/\pi\beta)$ is the function defined in 
(31). The function $\xi$ is a smooth function of $\lambda$ which is very
close to a straight line; it is sketched in figure 1 for different values
of the field $E$. \\
%
%
%
\indent
At first glance, the divergence of $B(\lambda)$ for $\lambda\to 0$ might
cause trouble in the integrals in (36). However, it follows from (37)
that $\mit\Psi(\lambda)\lambda^{-3/2}e^{1/2E\lambda}$ is locally integrable
at $\lambda=0$. Hence all integrals in (36) are well defined.

\section*{Acknowledgement}
One of the authors (CD) is grateful to the ``Studienstiftung des 
deutschen Volkes'' for financial support of this research.

\newpage
\begin{appendix}
%
\setcounter{equation}{0}
\renewcommand{\theequation}{\Alph{section}.\arabic{equation}}
\renewcommand{\thesubsection}{\Alph{section}.\Roman{subsection}}

\section{Appendix}
\boldmath
\subsection{Negativity of $\alpha(\lambda)$ for $\lambda\le E/\theta$}
\unboldmath
For $\lambda\le E/\theta$, the first integral in the definition of $\alpha(\lambda)$ 
(18) is a regular integral instead of a singular integral. Hence we may write this 
integral with the use of (15)
\begin{eqnarray*}
\int\limits_0^\infty\!d\mu\;\frac{{\cal N}(\mu)}{\mu(\lambda-\mu-E/\theta)} & = &
-\int\limits_0^\infty\!d\mu\;\frac{{\cal N}(\mu)}{\mu}\int\limits_0^\infty\!
dp\;\frac{p}{E}\;e^{(\lambda-\mu-E/\theta)p^2\!/2E} \\
& = & -\int\limits_0^\infty\!dp\,p\,
e^{(\lambda-E/\theta)p^2\!/2E - N(p)/E}
\end{eqnarray*}
Thus, combining terms in (18), we obtain for $\lambda\le E/\theta$
\begin{equation}
\alpha(\lambda) = -  \frac{\int\limits_{-\infty}^\infty\!dp\,p\,
e^{-\lambda p^2\!/2E+N(p)/E}\int\limits_p^\infty dq\,q\,
e^{(\lambda-E/\theta)q^2\!/2E - N(q)/E}}
{\int\limits_{-\infty}^\infty\!dp\,p\,e^{-\lambda p^2\!/2E+N(p)/E}} 
\end{equation}
Since $N(p)$ is an increasing function, the denominator of (A.1) is always
positive. Now we will prove that the numerator is positive too. For this purpose,
let us convert all integrals over regions with negative values of $p$ and $q$
into integrals over positive values. If we do so, we can write for the numerator
of (A.1)
\begin{eqnarray*}
 &   & \int\limits_0^\infty\!dp\,p\,e^{-\lambda p^2\!/2E+N(p)/E}
       \int\limits_p^\infty\!dq\,q\,e^{(\lambda-E/\theta)q^2\!/2E-N(q)/E} \\
 & - & \int\limits_0^\infty\!dp\,p\,e^{-\lambda p^2\!/2E+N(-p)/E}
       \int\limits_0^\infty\!dq\,q\,e^{(\lambda-E/\theta)q^2\!/2E-N(q)/E} \\
 & + & \int\limits_0^\infty\!dp\,p\,e^{-\lambda p^2\!/2E+N(-p)/E}
       \int\limits_0^p\!dq\,q\,e^{(\lambda-E/\theta)q^2\!/2E-N(-q)/E} 
\end{eqnarray*}
In the first term we write $\int_p^\infty dq=\int_0^\infty dq -\int_0^p dq$. 
Collecting terms with $\int_0^\infty dq$ and $\int_0^p dq$, the numerator of
(A.1) reads
\mathindent1.5cm
\begin{equation}
\int\limits_0^\infty\!dp\,p\,e^{-\lambda p^2\!/2E} \left\{
\int\limits_0^\infty\!dq\,q\,e^{(\lambda-E/\theta)q^2\!/2E}
\underbrace{\left(e^{N(p)/E-N(q)/E} - e^{N(-p)/E-N(q)/E}\right)}_{\textstyle 
>0\mbox{ for }p>0} \right. 
\end{equation}
\begin{displaymath}
\quad + \;\left. \int\limits_0^p\!dq\,q\,e^{(\lambda-E/\theta)q^2\!/2E} 
\underbrace{\left(e^{N(-p)/E-N(-q)/E} - e^{N(p)/E-N(q)/E}\right)}_{\textstyle 
<0\mbox{ for }p>q} 
 \right\} 
\end{displaymath}
\mathindent2cm
We can estimate the positive first term in the curly braces via 
$\int_0^\infty dq>\int_0^p dq$. Thus we arrive at the following lower estimate
for (A.2):
\begin{displaymath}
\int\limits_0^\infty\!dp\,p\,e^{-\lambda p^2\!/2E} 
\int\limits_0^p\!dq\,q\,e^{(\lambda-E/\theta)q^2\!/2E} 
\underbrace{\left(e^{N(-p)/E-N(-q)/E} - e^{N(-p)/E-N(q)/E}\right)}_{\textstyle 
>0\mbox{ for }q>0} 
\end{displaymath}
Since this lower estimate is positive, the numerator in (A.1) is positive, and 
consequently $\alpha(\lambda)$ is negative for $\lambda\le E/\theta$.

\boldmath
\subsection{Behaviour of $\alpha$ and $\beta$ for large $\lambda$}
\unboldmath
The index of our singular integral equation depends crucially on the values of
\begin{equation}
g(\lambda) = \frac{1}{2i}\ln\left(\frac{\alpha(\lambda)+\pi i\beta(\lambda)}
{\alpha(\lambda)-\pi i\beta(\lambda)}\right)
\end{equation}
for $\lambda\to0$ and $\lambda\to\infty$. Thus we need to determine the behaviour 
of the argument of the logarithm as $\lambda$ varies from zero to infinity.\par
First note that the absolute value of this argument is equal to unity for any value
of $\lambda$, because numerator and denominator are complex conjugates. 
Moreover, because of $\beta\equiv0$ for $\lambda\le E/\theta$, the argument is 
$+1$ for $\lambda\le E/\theta$ and with the choice $\ln(1)=0$ we have
\begin{displaymath}
g(\lambda)\equiv0 \quad\mbox{ for }\lambda\le E/\theta
\end{displaymath}
Now let us consider real and imaginary part of the argument of the logarithm
\begin{equation}
\mbox{Re}\left(\frac{\alpha+\pi i\beta}{\alpha-\pi i\beta}\right) =
\frac{\alpha^2-\pi^2\beta^2}{\alpha^2+\pi^2\beta^2} \quad\mbox{ and }\quad
\mbox{Im}\left(\frac{\alpha+\pi i\beta}{\alpha-\pi i\beta}\right) =
\frac{2\pi\alpha\beta}{\alpha^2+\pi^2\beta^2}
\end{equation}
We know from the previous section that $\alpha(E/\theta)$ is negative, and from
the definition (19) we know that $\beta(\lambda)<0$ for $\lambda>E/\theta$.
In consequence {\it the imaginary part is positive if $\lambda$ is slightly
greater than $E/\theta$.} \par
Moreover, from (A.4) we see that the real part cannot be positive whilst
the imaginary part is zero. In other words {\it the positive real axis cannot
be crossed for finite $\lambda$.} \par
In order to determine the behaviour for $\lambda\to\infty$, we use the 
lemma on the asymptotic behaviour of $\alpha$ and $\beta$ (see below). 
An immediate consequence of the lemma is that 
$\lim_{\lambda\to\infty}\beta/\alpha=0$ and $\alpha(\lambda)>0$ for large values
of $\lambda$. Hence {\it $(\alpha+\pi i\beta)/(\alpha-\pi i\beta)$ approaches
$+1$ with a negative imaginary part for $\lambda\to\infty$.} \par
Applying this result to our function $g(\lambda)$ defined in (A.3), we obtain
the limiting values
\begin{equation}
\lim_{\lambda\to0}g(\lambda)=0 \quad\mbox{ and }\quad
\lim_{\lambda\to\infty}g(\lambda)=\pi
\end{equation}
The logarithm may be expressed in terms of the arcus tangens function 
\begin{equation}
g(\lambda)=\mbox{arctan}\left(\frac{\pi\beta(\lambda)}{\alpha(\lambda)}\right)
=\mbox{arccot}\left(\frac{\alpha(\lambda)}{\pi\beta(\lambda)}\right)
\end{equation}
We must choose the branch of the arcus tangens, or arcus cotangens respectively,
in such a way that (A.6) is a continuous function of $\lambda$ and (A.5) is
satisfied. The resulting branches are sketched in figure 2. \\
%
%
%
{\bf Lemma:}
\begin{it}
As $\lambda$ approaches infinity, the asymptotic behaviour of $\alpha$ and
$\beta$ is given by
\begin{eqnarray}
\beta(\lambda) & \sim & \sqrt{E/2\pi}\,\nu(0)\, \lambda^{-3/2} \\[2mm]
\alpha(\lambda) & \sim & \sqrt{E/2\pi\lambda}\int\limits_{-\infty}^\infty\!
dp\,e^{-p^2\!/2\theta}\frac{\nu(p)}{\nu(0)}
=\frac{\sqrt{E/2\pi\lambda}}{M_0\,\nu(0)}
\end{eqnarray}
\end{it}
{\bf Proof:}
(A.7) is a consequence of (14) and the Tauberian theorems on the Laplace transform
(see \cite{Doetsch}, part 1), which may be applied because of ${\cal N}\ge 0$.
If we remember the definition $\beta(\lambda+E/\theta)=\Theta(\lambda)\,
{\cal N}(\lambda)/\lambda$, we see that (A.7) holds.\par
The asymptotic evaluation of $\alpha(\lambda)$ is a bit more intricate. 
Let us examine each term in the definition (18) separately: \par
The first term on the right hand side of (18) behaves like
\begin{equation}
\stackrel{\lambda\to\infty}{\sim}\frac{1}{\lambda-E/\theta}
\int\limits_0^\infty\!d\mu\,\frac{{\cal N}(\mu)}{\mu}=\frac{E}{\lambda-E/\theta}
\end{equation}
The denominator of the second term in (18) can be written as an integral over the 
positive real axis, which is easily evaluated asymptotically with {\sc Laplace}'s 
method (see \cite{Olver}, chapter 3)
\begin{equation}
\int\limits_0^\infty\!dp\,e^{-\lambda p^2\!/2E}
\underbrace{p\left(e^{N(p)/E}-e^{N(-p)/E}\right)}_{\stackrel{p\to 0}{\sim} 
2\nu(0)p^2\!/E}\stackrel{\lambda\to\infty}
{\sim}\nu(0)\sqrt{2\pi E}\;\lambda^{-3/2}
\end{equation}
The numerator of the second term on the right hand side of (18) can be 
evaluated as follows: after conversion of all integrals over the negative
real axis into integrals over the positive axis, we substitute 
$t=q^2\!/2E$ and $s'=p^2\!/2E$. If we then transform $s=s'-t$ and change
the order of integrations, we obtain
\mathindent0cm
\begin{displaymath}
E^2\int\limits_0^\infty\!ds\,e^{-\lambda s}\int\limits_0^\infty\!dt\,
e^{-Et/\theta} \left\{ 
e^{N\big(\sqrt{2E(s+t)}\big)/E-N\big(\sqrt{2Et}\big)/E} -
e^{N\big(-\sqrt{2E(s+t)}\big)/E-N\big(-\sqrt{2Et}\big)/E} \right\}
\end{displaymath}
\mathindent2cm
Again we may apply {\sc Laplace}'s method \cite{Olver}; we only need to 
determine the behaviour of the inner integral for $s\to 0$. The term
in curly braces behaves like
\begin{displaymath}
\Big\{ ...\Big\} \stackrel{s\to 0}{\sim} s\;
\frac{\nu\left(\sqrt{2Et}\right)+\nu\left(\sqrt{-2Et}\right)}
{\sqrt{2Et}}
\end{displaymath}
Thus the inner integral behaves like 
$s\int_{-\infty}^\infty\!dp\,e^{p^2\!/2\theta}\nu(p)/E$, and consequently
the numerator in (18) behaves like
\begin{equation}
\stackrel{\lambda\to\infty}{\sim}\frac{E}{\lambda^2}
\int\limits_{-\infty}^\infty\!dp\,e^{-p^2\!/2\theta}\nu(p)
\end{equation}
Collecting (A.9) - (A.11) and insertion into $\alpha(\lambda)$ according to
(18) yields (A.8). 
\end{appendix}


\begin{thebibliography}{99}
\bibitem{remark1} Remember that for positive kernels the eigenvalue with the 
largest absolute value is positive and the corresponding eigenfunction is 
positive too.
\bibitem{remark2} The simultanous normalization of $m$, $\tau$ and $\theta$
to unity can be achieved by transition to the dimensionless variables
\begin{displaymath}
\hat{p}:=p/\sqrt{\theta m} \quad\quad
\hat{x}:=x/\tau\sqrt{\theta} \quad\quad
\hat{E}:=E\tau/\sqrt{\theta m}
\end{displaymath}
\end{thebibliography}
\end{document}